\documentclass[aps,prd, onecolumn,superscriptaddress, 11pt]{revtex4} 
\usepackage{graphicx}
\usepackage{amssymb, amsmath, amsthm, amsfonts, bm, mathrsfs, wasysym, txfonts, bbm, enumerate}
\usepackage[usenames,dvipsnames]{xcolor}
\usepackage{epsfig}
\usepackage{graphicx}
\usepackage{amsfonts}
\usepackage{amssymb}
\usepackage{bm}
\usepackage{latexsym}
\usepackage{amsmath}
\usepackage{hyperref}
\usepackage{xcolor}
\usepackage{bbold}
\usepackage{slashed}
\begin{document}

\title{{\large Relativistic two-fluid hydrodynamics with quantized vorticity from \\ the
nonlinear Klein-Gordon equation} \vspace{3mm}}

\author{Chi Xiong}
\email[]{xiongchi@ntu.edu.sg}
\affiliation{Institute of Advanced Studies \& School of Physical and Mathematical Science, Nanyang Technological University, 639673 Singapore \vspace{5mm}}

\author{Kerson Huang}
\email[]{kerson@mit.edu. Deceased 1 September 2016.}
\affiliation{Physics Department, Massachusetts Institute of Technology, Cambridge,
MA, USA 02139 }

\begin{abstract}
\vspace{0.2cm}
\begin{center}
{\bf Abstract}
\end{center}

We consider a relativistic two-fluid model of superfluidity, in which the superfluid is described by an order parameter that is a complex scalar field satisfying the nonlinear Klein-Gordon equation (NLKG). The coupling to the normal fluid is introduced via a covariant current-current interaction, which results in the addition of an effective potential, whose imaginary part describes particle transfer between superfluid and normal fluid. Quantized vorticity arises in a class of singular solutions and the related vortex dynamics is incorporated in the modified NLKG, facilitating numerical analysis which is usually very complicated in the phenomenology of vortex filaments. The dual transformation to a string theory description (Kalb-Ramond) of quantum vorticity, the Magnus force and the mutual friction between quantized vortices and normal fluid are also studied.

\vspace{0.5cm}


\vspace{0.2cm}


\end{abstract}

\startpage{1}
\endpage{ }
\maketitle

\newpage

\section{Introduction}

Superfluidity is a macroscopic manifestation of quantum phase coherence, and
can be described in terms of of complex order parameter, which in the
relativistic domain satisfies a nonlinear Klein-Gordon equation (NLKG). We
regard the order parameter as the basic variable describing the superfluid,
and hydrodynamic variables, such as the superfluid density and velocity, as
derived quantities. Such a treatment not only conveys a more accurate physical
picture, but is also a highly efficient way to do numerical computations. 
In a previous paper \cite{Xiong14}, we consider a pure superfluid at absolute zero.
In this paper we extend the discussion to finite temperatures, where there is
also a normal fluid.

Being a manifestation of quantum phase coherence over macroscopic distances, 
superfluidity is best described in terms of of complex order parameter, which in 
the non-relativistic regime corresponds to a wave function satisfying the nonlinear 
Schr\"{o}dinger equation (NLSE). In the relativistic domain this is generalized to 
the NLKG. We regard the order parameter as the primary state variable of a superfluid, 
while hydrodynamic quantities, such as the superfluid density and velocity, are derived quantities. 
Such a view not only gives a more concrete physical picture, but, as shown in \cite{Xiong14},
also facilitates numerical analysis, especially in regard to quantized vorticity. 
We shall introduce the normal fluid in the same framework, and, to put it in historical perspective, 
start with a brief review of the two-fluid model \cite{Tisza38, Landau41}.

Shortly after superfluidity was discovered in liquid $^{4}$He below the
critical temperature of 2.8 K \cite{Kapitsa32}, as an apparent absence of
viscosity, Tisza \cite{Tisza38} suggested that the liquid be modeled as two
inter-penetrating liquids, one having ``super" qualities, and the other behaving
in a ``normal" fashion. Landau \cite{Landau41} made the model more concrete by
regarding the superfluid as the ground state of a quantum mechanical many-body
system, and the normal fluid as a system of quasiparticle excitations. The
mass density $\rho$ and mass current density $\bm{j}$ are split into
superfluid and normal fluid contributions. In the non-relativistic regime one
writes
\begin{align}
\rho &  =\rho_{s}+\rho_{n},\nonumber\\
\bm{j}  &  =\rho_{s}\bm{v}_{s}+\rho_{n}\bm{v}_{n},
\end{align}
where the subscripts $s$ and $n$ refer respectively to super and normal fluid,
with the conditon%
\begin{equation}
\nabla\times\bm{v}_{s}=0. \label{irrot}%
\end{equation}
The non-relativistic two-fluid hydrodynamics consists of phenomenological
equations based on conservation and thermodynamic laws \cite{Landau41}%
\cite{Khalatnikov65}:
\begin{align}
\frac{\partial\rho}{\partial t}+\nabla\cdot\bm{j}  &  =0,\nonumber\\
\left(  \frac{\partial}{\partial t}+\bm{v}_{s}\cdot\nabla\right)
\bm{v}_{s}  &  =-\nabla\mu,\nonumber\\
\frac{\partial j^{k}}{\partial t}+\partial_{j}\Pi^{jk}  &  =0,\nonumber\\
\frac{\partial S}{\partial t}+\nabla\cdot\left(  S\bm{v}_{n}\right)   &
=0, \label{2fluid}
\end{align}
where $\mu$ is the chemical potential, $S$ the entropy density, and $\Pi^{jk}$
is the energy-momentum tensor:
\begin{equation} \label{stress}
\Pi^{jk}=\rho_{s}v_{s}^{j}v_{s}^{k}+\rho_{n}v_{n}^{j}v_{n}^{k}+p\delta_{jk},
\end{equation}
where $p$ is the pressure. The second equation in (\ref{2fluid}) is the analog
of the Euler equation. At absolute zero $S$ vanishes identically, and the
equation for $j^{k}$ becomes the same as the Euler equation, and
(\ref{2fluid}) collapses to the first two equations describing a pure superfluid.

As thermodynamic functions, the quantities $S,\mu,p$ are to be specified in a
more detailed model. They can be calculated, for example, if the normal fluid
is modeled at low temperatures as a dilute gas of quasiparticles. Hill and
Roberts \cite{Hill77} introduced a special pressure term in $\mu$, in order to
describe the healing length, the characteristic distance within which the
superfluid density decreases to zero at a wall. Geurst \cite{Geurst80} has
given a general formulation of the two-fluid hydrodynamics in term of an
action principle, and a historical review. A relativistic action principle is
disussed by Lebedev and Khalatnikov \cite{Lebedev82}.

Even with a built-in healing length, however, the hydrodynamic equations fail
to describe one of the signature properties of a superfuid, namely, quantized
vorticity. In fact, the irrotational condition (\ref{irrot}) rules out
vorticity, and to accommodate that one has add it ``by hand", by writing
something like $\bm{v}_{s}=\nabla\alpha+\bm{b}$, where $\nabla
\times\bm{b}\neq0$, and go through another round of phenomenology for
$\bm{b}$. But all this still does not explain why the vorticity should be
quantized, not to mention the impracticality of numerical analysis.

All these difficulties in describing the superfluid are resolved by using a
complex order parameter
\begin{equation}
\Psi\left(  \bm{r},t\right)  =F\left(  \bm{r},t\right)  e^{i\beta
\left(  \bm{r},t\right)  },
\end{equation}
a non-relativistic wave function satisfying a nonlinear Schr\"{o}dinger
equation (NLSE). The superfluid velocity is related to the phase of the wave function through
\begin{equation}
\bm{v}_{s}=\frac{\hbar}{m}\nabla\beta,\label{NR}
\end{equation}
where $m$ is the mass scale in the NLSE. The healing length arises
automatically, since the superfluid density $\rho_{s}=mF^{2}$ goes to zero
continuously at a boundary. The Hill-Roberts pressure is just the ``quantum
pressure" arising naturally from the NLSE. The quantization of vorticity,
namely
\begin{equation}
{\displaystyle\oint_{C}}
d\bm{s} \cdot \bm{v}_{s}=\frac{2\pi\hbar}{m}n,\text{ }\left(  n=0,\pm
1,\pm2,\ldots\right)  ,
\end{equation}
where $C$ is a closed contour in space, follows from the fact that the phase
$\beta$ must be a continuous function \cite{Feynman}. In general $\nabla\times\bm{v}_{s}\neq0$, 
even though $\bm{v}_{s}$ is a gradient, because $\Psi$ can
develop zeros, thus rendering the space non-simply connected, i.e, admitting
closed contours that cannot be deformed to zero continuously. Another
advantage of the NLSE is that it can be handled numerically very efficiently.

Adopting the NLSE means that we regard the complex wave function $\Psi$ as the
fundamental variable, and the hydrodynamic quantities $\rho_{s},\bm{v}_{s}$ 
as derived ones. Thus, the first two equations in (\ref{2fluid}) are
replaced by and implied by the NLSE.

To include the normal fluid in the NLSE, we need to introduce four new degrees
of freedom associated with $\rho_{n},\bm{v}_{n}$. Bogolubov apparently was
the first to suggest, in an unpublished note \cite{Shygoin09}, the
introduction of gauge-like potentials $\varphi,\bm{A}$ via the transformation
\begin{align}
\frac{\partial}{\partial t}  &  \rightarrow\nabla-i\varphi\nonumber\\
\nabla &  \rightarrow\nabla-i\bm{A}%
\end{align}
This is done in order to couple the new degrees of freedom to the phase of the
wave function; the system is of course not locally gauge-invariant. (It had
better not be, for otherwise the above would have no physical effect.) Coste
\cite{Coste98} shows how one can relate $\varphi,\bm{A}$ to $\rho
_{n},\bm{v}_{n}$ through considerations based on Galilean invariance. To
obtain the equations of motion for $\rho_{n},\bm{v}_{n}$, Coste uses a
hybrid variational principle involving $\Psi,\rho_{n},\bm{v}_{n}$. 

This paper is organized in the following manner. After a brief description of
the NLKG at absolute zero, we extend it to finite temperature by introducing
couplings to the normal fluid, based on Lorentz covariance. We show that the
couplings can be expressed in terms of an additional nonlinear potential that
has both a real and imaginary parts, and discuss its non-relativistic limit.
We display quantized vorticity by transforming the scalar field theory to a global
string theory. Magnus force and mutual friction are extracted in some simple examples.

\section{ NLKG (Nonlinear Klein-Gordon equation)}

In this section we use units in which $\hbar=c=1$. 
Consider a complex scalar field $\phi (\bm{x}, t)$, which can be written 
in the phase representation as
\begin{equation}
\phi (\bm{x}, t) =F(\bm{x}, t) ~ e^{i\sigma(\bm{x}, t)}.
\end{equation}
It serves as order parameter for superfluidity through the dynamics of the
phase $\sigma(\bm{x}, t)$. The physical superfluid velocity $\bm{v}_{s}$ is related
to the 4-vector%
\begin{equation} \label{v_mu}
v^{\mu}=\partial^{\mu}\sigma
\end{equation}
through%
\begin{equation} \label{v_s}
\bm{v}_{s}=\frac{\bm{\nabla} \sigma}{\omega},
\end{equation}
Here, $\omega$ is a frequency given by the time component of $v^{\mu}:$%
\begin{equation} \label{omega}
\omega\equiv\partial^{0}\sigma,
\end{equation}
which ensures $|\bm{v}_{s}|/c <1$. The Langrangian density is given by
\begin{equation}
\mathcal{L}_{0}=g^{\mu\nu}\partial_{\mu}\phi^{\ast}\partial_{\nu}\phi+V,
\end{equation}
where $g^{\mu\nu}$ is the metric tensor. The potential $V$ depends only on
$\phi^{\ast}\phi$, and $V^{\prime}\equiv dV/d\left(  \phi^{\ast}\phi\right)
$. The action
\begin{equation}
S_{0}=-\int d^{4}x\sqrt{-g}\mathcal{L}_{0}\mathcal{=}-\int d^{4}x\sqrt
{-g}\left(  g^{\mu\nu}\partial_{\mu}\phi^{\ast}\partial_{\nu}\phi+V\right)
\end{equation}
leads to the NLKG
\begin{equation}
\left(  \square-V^{\prime}\right)  \phi=0,
\end{equation}
where $\square\phi\equiv\frac{1}{\sqrt{-g}}\partial_{\mu}\left(  \sqrt
{-g}g^{\mu\nu}\partial_{\nu}\phi\right)$. Some examples of NLKG in rotating blackhole 
backgrounds are given in \cite{Good14}.
In the phase representation, the real and imaginary parts of the NLKG give rise to two
hydrodynamic-like equations
\begin{align} \label{hydro}
\left(  \square-V^{\prime}\right)  F-F\nabla^{\mu}\sigma\nabla_{\mu}\sigma &
=0,\nonumber\\
2\nabla^{\mu}F\nabla_{\mu}\sigma+F\nabla^{\mu}\nabla_{\mu}\sigma &  =0.
\end{align}
The first is the analog of the Euler equation, and the second is the
continuity equation $\ \nabla_{\mu}j_{0}^{\mu}=0,$ where $\nabla_{\mu}$
denotes the covariant derivative, with
\begin{equation}
j_{0}^{\mu}=\frac{1}{2i}\left(  \phi^{\ast}\partial^{\mu}\phi-\phi
\partial^{\mu}\phi^{\ast}\right)  =F^{2}\partial^{\mu}\sigma
\end{equation}
This is a number current density, corresponding to the conservation of charge
$Q=N-\bar{N}$, where $N,$ $\bar{N}$ are respectively the particle and
antiparticle number. Unlike the non-relativistic case, the density
$j_{0}^{(0)}$ is not positive definite.

If $\sigma$ were a continuous function, we would have $\partial^{\mu}v^{\nu
}-\partial^{\nu}v^{\mu}=\left(  \partial^{\mu}\partial^{\nu}-\partial^{\nu
}\partial^{\mu}\right)  \sigma=0$. But $\sigma$ is a phase angle, and only
continuous modulo $2\pi$. The derivatives $\partial^{\mu},\partial^{\nu}$ do
not commute when operating on $\sigma$, and this is the origin of quantized
vorticity. Specifically, there is a class of singular solutions in which the
modulus $F$ has zeros along a space curve, the vortex line, and $
{\displaystyle\oint_{C}}\bm{\nabla}\sigma\cdot d\bm{s=}2\pi n$ 
along any closed circuit $C$ encircling the vortex line, where $n$ is an integer.

Without going into details, we comment on the fact that the first equation in
(\ref{hydro}) can be rewritten in Euler form as an equation for $d\bm{v}/dt$, 
which contains a ``quantum pressure". This pressure naturally vanishes on
boundaries where $F$ goes to zero, with a healing length.There is no need for
the ``Hill-Roberts pressure" \cite{Hill77}, which is introduced by hand.

\section{NLKG with normal fluid: the effective potential}

In the following, we consider Minkowski spacetime with $\eta^{\mu\nu}=$
diag$( -1,1,1,1) $.  At finite temperatures, new degrees freedom arise,
associated with the normal-fluid velocity field $\bm{v}_{n}$. We represent
the new degrees of freedom covariantly by a 4-vector $w^{\mu}$, and, following
Lebedev and Khalatnikov \cite{Lebedev82}, write it in a Clebsch representation
of the form
\begin{equation}
w^{\mu}=\partial^{\mu}\alpha+\xi\partial^{\mu}\chi
\end{equation}
where $\alpha,\xi,\chi,$ are regarded as independent variables, which will be
related to physical normal fluid properties. 
We can construct three Lorentz invariants From $v^{\mu}$ and $w^{\mu}$:
\begin{equation}
I_{1}=\frac{F^{2}}{2}v^{\mu}v_{\mu}, ~~~I_{2}=F^{2}v^{\mu}w_{\mu
}, ~~~I_{3}=\frac{F^{2}}{2}w^{\mu}w_{\mu}
\end{equation}
and generalize the  zero-temperature Lagrangian density $\mathcal{L}_0$ to
\begin{equation}
\mathcal{L}=\mathcal{L}_{0}+f\left(  I_{1},I_{2},I_{3}\right)
\label{lagrangian}%
\end{equation}
where $f$ is a function with derivatives denoted by
\begin{equation}
f_{n}^{\prime}=\frac{\partial f}{\partial I_{n}}
\end{equation}
The equations of motion are obtained by varying the new action with respect to
$F,\sigma,\alpha,\xi,\chi:$
\begin{align} 
\left[  \square-V^{\prime}-\left(  1+f_{1}^{\prime}\right)  v_{\mu}v^{\mu
}-2f_{2}^{\prime}w_{\mu}v^{\mu}-f_{3}^{\prime}w_{\mu}w^{\mu}\right]  F  &
=0\nonumber\\
\partial_{\mu}j^{\mu}  & =0\nonumber\\
\partial_{\mu}s^{\mu}  & =0\nonumber\\
s^{\mu}\partial_{\mu}\xi & =0\nonumber\\
s^{\mu}\partial_{\mu}\chi & =0\label{list}
\end{align}
where two current densities $j^\mu$ and $s^\mu$ are defined as
\begin{align}
j^{\mu}  & \equiv F^{2}\left[  \left(  2+f_{1}^{\prime}\right)  v^{\mu}%
+f_{2}^{\prime}w^{\mu}\right]  \nonumber\\
s^{\mu}  &  \equiv F^{2}\left(  f_{2}^{\prime}v^{\mu}+f_{3}^{\prime}w^{\mu}\right).
\label{currents}
\end{align}
These are conserved current densities, identified respectively with the number current density 
and the entropy current density. The former reduces to $j_{0}^{\mu}$ when $f\equiv0$, and 
the latter is present only at nonzero temperatures.
The first two equations of motion in (\ref{list}) can be combined to give a new NLKG:
\begin{equation} \label{NLKGeffW}
\left(  \square-V^{\prime}-W\right)  \phi=0
\end{equation}
where $W$ is an effective potential obtained by plugging $\phi = F e^{ i \sigma}$ 
into (\ref{NLKGeffW}) and then comparing with the first two equations of (\ref{list})
\begin{eqnarray} \label{Eff_W}
W &=& f_{1}^{\prime}v_{\mu}v^{\mu}+2f_{2}^{\prime}w_{\mu}v^{\mu} +f_{3}^{\prime}w_{\mu}w^{\mu}+ i ~ F^{-2}\partial_{\mu}\left(F^{2}v^{\mu}\right) \cr
&=&  f_{1}^{\prime}v_{\mu}v^{\mu}+2f_{2}^{\prime}w_{\mu}v^{\mu}%
+f_{3}^{\prime}w_{\mu}w^{\mu} -i \left[v^{\mu} \partial_\mu  f_1^{\prime} + F^{-2} \partial_{\mu} (F^2 w^\mu f_2^{\prime})\right]/(2+ f_1^{\prime}).
\end{eqnarray}
Note that the effective potential $W$ has both real and imaginary parts, 
indicating that the NLKG for the superfluid is dissipative: there is particle transfer 
between superfluid and normal fluid.
Assuming that $\alpha,\beta,\xi$ are continuous functions, we have
$\partial^{\mu}w^{\nu}-\partial^{\nu}w^{\mu}=\partial^{\mu}\xi\partial^{\nu
}\beta-\partial^{\nu}\xi\partial^{\mu}\beta.$ Multiplying by $s_{\mu}$ yields%
\begin{equation}
s_{\mu}\left(  \partial^{\mu}w^{\nu}-\partial^{\nu}w^{\mu}\right)  =0
\end{equation}
We will show that the imaginary part of the effective potential $W$ is important in
determining the mutual friction coefficients, which describe the coupling between quantized
vortices and the normal fluid. 

\section{The normal-fluid}

We represent the time and spatial components of current densities $j^{\mu}, s^{\mu}$ as
\begin{eqnarray} \label{s}
j^{\mu} &  =\left(  \rho,\bm{j}\right)  \nonumber\\
s^{\mu} &  =\left(  s, s \bm{v}_{n}\right)  
\end{eqnarray}
The first relation defines total number density $\rho$ and current
$\bm{j,}$ with number meaning $N-\bar{N},$ the difference between
particle and antiparticle number. The second defines the entropy density $s$
and the normal-fluid velocity $\bm{v}_{n}.$ (The definition of $\rho$
agrees with Lebedev and Khalatnikov \cite{Lebedev82}, but differs from our
earlier paper \cite{Xiong14}, in which the density corresponds to $\rho
\sqrt{1-v_{s}^{2}}$ in the present convention). The conservation laws can
be put in the form
\begin{align}
\frac{\partial\rho}{\partial t}+\nabla\cdot\bm{j} &  =0\nonumber\\
\frac{\partial s}{\partial t}+\nabla\cdot\left(  s\bm{v}_{n}\right)   &
=0
\end{align}
The second equation of (\ref{currents}) can be rewritten as
\begin{equation}
w^{\mu}=\frac{s^{\mu}}{F^{2}f_{3}^{\prime}}-\frac{f_{2}^{\prime}}
{f_{3}^{\prime}}v^{\mu}
\end{equation}
Substitution into the first equation of (\ref{currents}) gives
\begin{equation}
j^{\mu}= F^{2}\left(  2+f_{1}^{\prime}-\frac{f_{2}^{\prime2}}
{f_{3}^{\prime}}\right)  v^{\mu}+\frac{f_{2}^{\prime}}{f_{3}^{\prime}}s^{\mu}
\end{equation}
whose components are given by
\begin{align}
\rho &  = F^{2}\left(  2+f_{1}^{\prime}-\frac{f_{2}^{\prime2}}
{f_{3}^{\prime}}\right)  \omega+\frac{f_{2}^{\prime}}{f_{3}^{\prime}
}s \nonumber\\
\bm{j}  &  = F^{2}\left(  2+f_{1}^{\prime}-\frac{f_{2}^{\prime
2}}{f_{3}^{\prime}}\right)  \omega\bm{v}_{s}+\frac{f_{2}^{\prime}}
{f_{3}^{\prime}}s\bm{v}_{n}
\end{align}
We make the identification
\begin{align}
\rho_{s}  &  =\omega F^{2}\left(  2+f_{1}^{\prime}-\frac
{f_{2}^{\prime2}}{f_{3}^{\prime}}\right) \nonumber\\
\rho_{n}  &  =\frac{f_{2}^{\prime}}{f_{3}^{\prime}}s \label{rhos}
\end{align}
such as to give
\begin{align}
\rho &  =\rho_{s}+\rho_{n}\nonumber\\
\bm{j}  &  =\rho_{s}\bm{v}_{s}+\rho_{n}\bm{v}_{n}
\end{align}
We can write, in manifestly covariant form,
\begin{equation}
j^{\mu}=\frac{\rho_{s}}{\omega}v^{\mu}+\frac{\rho_{n}}{s}s^{\mu}
\end{equation}
The energy-momentum tensor of the superfluid is given by
\begin{align}
T^{\mu\nu}  &  =\frac{\partial\mathcal{L}}{\partial\left(  \partial_{\mu
}F\right)  }\partial^{\nu}F+\frac{\partial\mathcal{L}}{\partial\left(
\partial_{\mu}\sigma\right)  }\partial^{\nu}\sigma+\frac{\partial\mathcal{L}
}{\partial\left(  \partial_{\mu}\alpha\right)  }\partial^{\nu}\alpha
+\frac{\partial\mathcal{L}}{\partial\left(  \partial_{\mu}\beta\right)
}\partial^{\nu}\beta-g^{\mu\nu}\mathcal{L}\nonumber\\
&  =2\partial^{\mu}F\partial^{\nu}F+j^{\mu}v^{\nu}+s^{\mu}w^{\nu}-g^{\mu\nu
}\mathcal{L}
\end{align}
Using (\ref{currents}), we obtain the symmetric form
\begin{equation}
T^{\mu\nu}=2\partial^{\mu}F\partial^{\nu}F+\frac{\rho_{s}}{\omega}v^{\mu
}v^{\nu}+\frac{\rho_n}{\tilde{s}}s^{\mu}s^{\nu} -g^{\mu\nu}\mathcal{L}
\end{equation}
where $\tilde{s} \equiv \rho_n F^{2}f_{3}^{\prime}$. 
The spatial components give the stress-energy tensor
\begin{equation} 
T^{jk}=2\partial^{j}F\partial^{j}F+\omega \, \rho_{s}v_{s}^{j}v_{s}^{k}
+ (s^{2}/\tilde{s}) \, \rho_n \, v_{n}^{j}v_{n}^{k}-\delta^{jk}\mathcal{L}
\end{equation}
where we have used (\ref{v_s}) and (\ref{s}). The coefficient of $v_{n}^{j}v_{n}^{k}$ 
should be $\omega\rho_{n}$, as suggested by comparison with (\ref{stress}). Thus we have the relation
\begin{equation}
s^{2} = \omega \tilde{s}   \label{f3}
\end{equation}
Using (\ref{f3}) and (\ref{rhos}), we can now express the parameters
$f_{n}^{\prime}$ in terms of observable normal-fluid properties:
\begin{align}
f_{1}^{\prime} &  =\frac{\rho}{\rho_{0}} -2  \nonumber\\
f_{2}^{\prime} &  =\frac{s}{\rho_{0}}\nonumber\\
f_{3}^{\prime} &  =\frac{s^{2}}{\rho_{0}\rho_{n}}
\end{align}
where $\rho_{0}$ is the density at absolute zero temperature:
\begin{equation}
\rho_{0}\equiv F^{2}\omega
\end{equation}
The NLKG describes the superfluid and it coupling to the normal fluid, whose
dynamics requires separate treatment. A general hydrodynamic treatment based on
conservation laws and thermodynamics has been given by Lebedev and Khalatnikov
\cite{Lebedev82} and by Carter and Khalatnikov \cite{Carter92}. A general action
priniple has been discussed by Geurst \cite{Geurst80}. A more detailed
description of the normal fluid will depend on the specific model. In this
respect, a relativistic ideal gas model of the normal fluid may be found in
\cite{Carter95}, and a treatment based on quantum field theory of the scalar
field is given in \cite{Alford13}.

Note that the physical meaning of the currents and densities depend on the reference frames and we compare our results with \cite{Carter95}. We consider the following Lorentz scalars:
\begin{eqnarray} \label{LS}
v^{\mu} v_{\mu} &=& - c^2 \mu^2 \cr
s^{\mu} s_{\mu} &=& - c^2 s^2 \cr
v^{\mu} s_{\mu} &=& - c^2 y^2
\end{eqnarray}
Our choices for $v^{\mu}$ and $S^{\mu}$ are
\begin{eqnarray}
v^{\mu} &=& ( \tilde{\mu}, \nabla \sigma) = \tilde{\mu} (1,  {\bm v}_s) \cr
s^{\mu} &=&  \tilde{s} (1, {\bm v}_n).
\end{eqnarray}
(Here we changed some notations for comparison purpose. We defined before in the equations (\ref{v_mu}), ({\ref{v_s}) and (\ref{omega})
\begin{equation}
v^{\mu}=\partial^{\mu}\sigma, ~~ \bm{v}_{s}=\frac{\bm{v}}{\omega} = \frac{\nabla \sigma}{\omega}, ~~\omega\equiv\partial^{0}\sigma,
\end{equation}
and we have changed $\omega$ to $\tilde{\mu}$, and $s$ to $\tilde{s}$ respectively.)
Note that the velocities ${\bm v}_s$ and ${\bm v}_n$ are defined in the lab frame.
From (\ref{LS}) we obtain
\begin{eqnarray}
\tilde{\mu} &=& \gamma_s \mu, ~~ \gamma_s \equiv \frac{1}{\sqrt{1 - \frac{{\bm v}_s^2}{c^2}}},  \cr
\tilde{s} &=& \gamma_n s, ~~ \gamma_n \equiv \frac{1}{\sqrt{1 - \frac{{\bm v}_n^2}{c^2}}}.
\end{eqnarray}
While $\mu, s$ are Lorentz scalars, $\tilde{\mu}, \tilde{s}$ are not due to the $\gamma$-factors in the above equations. The crossing term yields
\begin{equation} \label{vsy}
v^{\mu} s_{\mu} \equiv - c^2 y^2 = - c^2 ~\tilde{\mu} \,\tilde{s}~ \left( 1 - \frac{{\bm v}_s \cdot {\bm v}_n}{c^2} \right) =  - c^2 ~\mu\,s~\gamma_n \gamma_s ~ \left( 1 - \frac{{\bm v}_s \cdot {\bm v}_n}{c^2} \right).
\end{equation}
We define a relative velocity, ${\bm v}_{ns}$, between the superfluid velocity ${\bm v}_s$ and the normal velocity ${\bm v}_n$ according to the relativistic velocity-addition formula
\begin{equation}
{\bm v}_{ns} = \frac{{\bm v}_{n} - {\bm v}_{s}}{ 1 - \frac{{\bm v}_s \cdot {\bm v}_n}{c^2}},
\end{equation}
and then find
\begin{equation}
1 - \frac{{\bm v}_{ns}^2}{c^2} = \frac{1}{\gamma_n^2 \gamma_s^2  \left( 1 - \frac{{\bm v}_s \cdot {\bm v}_n}{c^2} \right)^2 }. 
\end{equation}
Plugging into (\ref{vsy}) we obtain
\begin{equation}
\frac{{\bm v}_{ns}^2}{c^2}  = 1 - \frac{\mu^2 s^2}{y^4}
\end{equation}
which is exactly the relative translation speed between the ``normal" and ``superfluid" frames used in Ref. \cite{Carter95}. One can identify the relativistic generalization of the mass densities of the superfluid and normal fluids, $\hat{\rho}_n$ and $\hat{\rho}_s$, by considering the decomposition of the stress-energy tensor $T^{\mu\nu}$. The advantage of finding $\hat{\rho}_n$ and $\hat{\rho}_s$ based on the decomposition of $T^{\mu\nu}$ is that it avoids the use of ``rest mass", more precisely, the question on which frame should be considered as the rest frame of the fluid element. What is needed is the relative translation velocity between the normal fluid and the superfluid frames \cite{Carter95}. Note that 
\begin{equation}
y^2 = \mu_n \, s = \mu \, s_s, ~~\textrm{where}~\mu_n \equiv \mu/\sqrt{1- \frac{v_{ns}^2}{c^2}}, ~s_s \equiv s /\sqrt{1- \frac{v_{ns}^2}{c^2}}
\end{equation}
where $s_s$ is the entropy density defined in the superfluid frame and the $\mu_n$ is the chemical potential defined in the normal fluid frame. The stress-energy tensor becomes
\begin{align}
T^{\mu\nu}  & =  2\partial^{\mu}F\partial^{\nu}F+\frac{\hat{\rho}_{s}}{\mu_n^2}v^{\mu
}v^{\nu}+\frac{\hat{\rho}_n}{s_s^2}s^{\mu}s^{\nu}-g^{\mu\nu}\mathcal{L}\cr
\hat{\rho}_{s}  & \equiv   \mu_n^2 \, F^2 \left[  2+f_{1}^{\prime}-\frac{\left(  f_{2}^{\prime}\right)^{2}}{f_{3}^{\prime}}\right]  \cr
\hat{\rho}_{n}  & \equiv   s_s^2 \, \frac{1}{F^2 f_3^{'}}.
\end{align}
It is not hard to check that these are consistent with the results in Ref.\cite{Carter95}. In fact the explicit correspondences are 
\begin{eqnarray}
\mathcal{A} &=& - \frac{\omega}{s} \frac{\rho_n}{\rho_s} \cr
\mathcal{B} &=& \frac{\omega}{\rho_s} \cr
\mathcal{C} &=& \frac{\omega \rho_n}{s^2} \left( 1 + \frac{\rho_n}{\rho_s} \right)
\end{eqnarray}
where $\mathcal{A}, \mathcal{B}, \mathcal{C}$ are quantities defined in \cite{Carter95}
\begin{eqnarray}
w^{\mu} &=& \mathcal{C} s^{\mu} + \mathcal{A} j^{\mu}, \cr
v^{\mu} &=& \mathcal{A} s^{\mu} + \mathcal{B} j^{\mu},
\end{eqnarray}


\section{The non-relativistic limit: modified NLSE}

A solution $\phi$ of the NLKG contains both positive and negative frequencies.
It approaches the nonrelativistic limit when one sign (say, positive) becomes
dominant. Formally, we write
\begin{equation}
\phi\underset{c \rightarrow\infty}{\longrightarrow}\Psi e^{-i\left(
mc^{2}/\hbar\right)  t}
\end{equation}
where $m$ is a large mass scale. The nonrelativistic wave function $\Psi$ can
be represented in the form
\begin{equation}
\Psi=\sqrt{\rho}e^{i\beta}
\end{equation}
where $\rho$ is the non-relativistic superfluid density, and
\begin{equation}
\bm{v}_{s}=\frac{\hbar}{m}\nabla\beta
\end{equation}
is the non-relativistic superfluid velocity. The nonrelativistic phase $\beta$
is related to the phase $\sigma$ of the relativistic scalar field $\phi$
through
\begin{equation}
\dot{\beta}= \dot{\sigma} + \frac{mc^{2}}{\hbar}, ~~~~~\nabla \beta = \nabla \sigma.
\end{equation}
The wave function $\Psi$ satisfies an NLSE (nonlinear Schr\"{o}dinger
equation) (see \cite{Xiong14} for details). To derive it, it is easier to start from the Lagrangian. Let
$\mathcal{L}_{0}$ be the nonrelativistic Lagrangian density at absolute zero,
which leads to an NLSE with cubic nonlinearity. We show how the normal fluid
may be introduced, following Coste \cite{Coste98}, but reformulated from our
point of view.

Let the superfluid density and current density be denoted respectively by
$\rho$ and $\bm{j}=\rho\bm{v}_{s}$ . The degrees of freedom
$\rho_{n}$, $\bm{v}_{n}$ associated with the normal fluid can be
introduced via gauge-like potentials $\varphi,\bm{A,}$ through the
transformation $\frac{\partial}{\partial t}\rightarrow\nabla-i\varphi,$
$\nabla\rightarrow\nabla-i\bm{A}$. This method was apparently first
suggested by Bogoliubov in an unpublished note \cite{Shygoin09}.The Lagrangian
density at finite temperatures is
\begin{equation}
\mathcal{L}=\mathcal{L}_{0}+\rho\varphi-\bm{j}\cdot\bm{A}+\frac{1}
{2}\rho A^{2}
\end{equation}
This is not locally gauge-invariant, (and had better not be, for otherwise the
gauge transformation would have no physical effect.) The term $\frac{1}{2}\rho
A^{2}$, while crucial for local gauge invariance, is irrelevant here, and will
be dropped. Using arguments based on Galillean covariance, Coste
\cite{Coste98} writes
\begin{align}
\bm{A}  &  =\alpha\left(  \bm{v}_{s}-\bm{v}_{n}\right) \nonumber\\
\varphi &  =\bm{v}_{n}\cdot\bm{A}
\end{align}
where $\alpha$ is a scalar function, and $\bm{v}_{n}$ is the normal-fluid
velocity. The equations of motion are then obtained through the action
principle. We omit details and just cite the final result. Assuming an original NLSE 
with quartic nonlinearity, we obtain a modified equation
\begin{align}
i\frac{\partial\Psi}{\partial t}  &  =-\frac{1}{2}\nabla^{2}\Psi+\left(
|\Psi|^{2}-1+U\right)  \Psi\nonumber\\
U  &  =-\frac{1}{2}\left(  \bm{v}_{n}-\bm{v}_{s}\right)  ^{2}
\frac{\partial\rho_{n}}{\partial\rho}-\frac{i}{2\rho}\nabla\cdot\left[
\rho_{n}\left(  \bm{v}_{n}-\bm{v}_{s}\right)  \right]
\end{align}
where $\hbar=m=1,$ and all coupling parameters have been scaled to unity. The
normal fluid enters via the effective potential $U$, which vanishes at
absolute zero. The real part of $U$ contributes to the phase change of $\Psi$,
and thus to superfluid flow. The imaginary part contributes to $\dot{\Psi}$,
rendering $\int d^{3}x|\Psi|^{2}$ non-conserved, signifying particle transfer
between superfluid and normal fluid.

\section{From phenomenology of quantized vorticity to NLKG formulation}

A phenomenological treatment of quantized vorticity in the non-relativistic
domain was pioneered by Schwarz \cite{Schwarz} in the non-relativistic domain,
based on the following physical picture (see Appendix A for notations). A vortex configuration is
characterized by a space curve called the vortex line, described by the
position vector $\bm{s}(\xi,t)$, where $\xi$ is a parameter that runs
along the line. The vortex line may be made up of disjoint closed loops, and
curves that terminate on boundaries. The parameter $\xi$ run through all of the
components according to some convention. The superfluid density vanishes on
the vortex line with a characteristic healing length. We can picture the
vortex line as a tube with effective radius $a_{0}$ of the order of the
healing length. This core size is supposed to be much smaller than any other
length in the theory. When we refer a point on the vortex line, we mean some
point within the core. Let $\bm{s}^{\prime}\equiv\partial\bm{s}/\partial\xi$. 
The triad $\bm{s}, \bm{s}^{\prime},\bm{s}^{\prime\prime}$ gives a local
orthogonal coordinate system. The local radius of curvature is given by
$R=\left\vert \bm{s}^{\prime\prime}\right\vert $ $^{-1}.$

The superfluid velocity is determined up to a potential flow by the equation
\begin{equation}
\nabla\times\bm{v}_{s}\bm{=\kappa}
\end{equation}
where $\bm{\kappa} (\bm{r},t)  $ is the vorticity density, which is
nonvanishing only on the vortex line:
\begin{equation}
\bm{\kappa} (\bm{r}, t)  =\kappa_{0}\int d\bm{s} \, \delta (\bm{r}-\bm{s}(\xi, t) )
\end{equation}
We decompose the superfluid velocity into an irrotational part $\bm{v}_{0}$, 
and a rotational part $\bm{b}:$
\begin{align}
\bm{v}_{s}  &  \bm{=}\bm{v}_{0}+\bm{b}\nonumber\\
\nabla\times\bm{v}_{0}  &  =\nabla\cdot\bm{b}=0\nonumber\\
\nabla\times\bm{b}  &  \bm{=\kappa} \label{decomp}%
\end{align}
The velocity field $\bm{b}$ is like a magnetic field produced by the
current density $\bm{\kappa}$, and is given by the Biot-Savart law
\begin{equation}
\bm{b}(\bm{r},t)=\frac{\kappa_{0}}{4\pi}\int\frac{\left(
\bm{s}_{1}-\bm{r}\right)  \times d\bm{s}_{1}}{\left\vert
\bm{s}_{1}- \bm{r} \right\vert ^{3}} \label{biot-savart}
\end{equation}
where $\bm{s}_{1}$ is a particular point on the vortex line.

The velocity of the vortex line at any point is influence by the shape of the
entire vortex line. In a ``local inducting approximation" (LIA), one considers only the effects
from the immediate neighborhood of the point. In this case, this the local
velocity is the translational velocity of an osculating vortex ring at that
point, which is normal to the plane of the vortex ring, and approximately
inversely proportional to its radius of curvature $R$. For a vortex line at
absolute zero, this leads to the equation
\begin{align} \label{beta}
\bm{\dot{s}}_{0}  &  =\beta\bm{s}^{\prime}\times\bm{s}^{\prime\prime}+\bm{v}_{s}\nonumber\\
\beta &  =\frac{\kappa_{0}}{4\pi}\ln\left(  \frac{c_{0}\bar{R}}{a_{0}}\right)
\end{align}
where $\bm{\dot{s}}_{0}\bm{\equiv\partial s}_{0}\bm{/}\partial t$,
$\bar{R}$ is the average radius of curvature, and $c_{0}$ is a constant of
order unity.

At finite temperatures, there is a normal fluid, which exerts a dissipative
force per unit length\ $\bm{f}_{D}$\ on the vortex line. It can be fit
phenomenologically by the formula
\begin{equation}
\frac{{\bm f}_{D}}{\rho_s \kappa_0} = - \alpha {\bm s}' \times \left[ {\bm s}' \times ( {\bm v}_{ns} - {\bm v}_{sl}) \right] - \alpha'  {\bm s}' \times ( {\bm v}_{ns} - {\bm  v}_{sl})
\end{equation}
where $\bm{v}_{ns}=\bm{v}_{n}-\bm{v}_{s}$, and $\alpha
,\alpha^{\prime}$ are temperature-dependent parameters.
The vortex line experiences a Magnus force per unit length
$\bm{f}_{M},$ when the vortex line velocity $\bm{v}_{L}\left(  \xi,t\right)
\equiv\bm{\dot{s}}\left(  \xi,t\right)  $ is different from the local
superfluid velocity $\bm{v}_{sl}\left(  \xi,t\right)  \equiv\bm{v}_{s}\left(\bm{s}\left(  \xi,t\right)  ,t\right) $:
\begin{equation}
\frac{\bm{f}_{M}}{\rho_{s}\kappa_{0}}=\bm{s}^{\prime}
\times\left(  \bm{v}_{L}-\bm{v}_{sl}\right)
\end{equation}
The phenomenological equations give physical insight, but for actual computations it is simpler to use the NLKG directly. As shown in \cite{Xiong14}, complex phenomena such as vortex formation and reconnection can be exhibited in numerical solutions of the NLKG. When quantum vorticity appears, the phase $\sigma$ of the complex field cannot be smooth everywhere, hence $\nabla_{\mu} v_{\nu} - \nabla_{\nu} v_{\mu} \ne 0$. From numerical calculations, the phase $\sigma$ has ambiguity and the modulus $F$ vanishes at the locations of vortices. 
To be consistent with the NLKG with effective potential describing the normal fluid effects, the variational principle should be applied to the Lagrangian
\begin{equation} \label{Nvp}
\mathcal{L} (f_{\mu},  v_{\mu}, w_{\mu}) = \mathcal{L} (F, \nabla_{\mu} F,  \nabla_\mu \sigma,  \nabla_\mu\alpha, \zeta,  \nabla_\mu\beta)
\end{equation}
where $f_{\mu} \equiv \nabla_{\mu} F$ and $\nabla \sigma$ is generally NOT a smooth function.  Note that one can split the Lagrangian into
\begin{equation}
\mathcal{L} = \mathcal{L}^0_{\textrm{\tiny{NLKG}}} +  \mathcal{L}^{\textrm{T}}
\end{equation}
where $\mathcal{L}^0_{\textrm{\tiny{NLKG}}}$ is similar to the zero-temperature cases
\begin{equation}
\mathcal{L}^0_{\textrm{\tiny{NLKG}}} = - g^{\mu\nu} \partial_{\mu} F \partial_{\nu} F - F^2 g^{\mu\nu} \partial_{\mu} \sigma \partial_{\nu} \sigma - V (F^2),
\end{equation}
while $\mathcal{L}^{\textrm{T}}$ includes finite-temperature effects. Following similar notations in \cite{Lebedev82}, we rewrite
\begin{eqnarray} \label{new_vw}
v_{\mu} &=& \nabla_\mu \sigma \equiv \nabla_{\mu} \varphi + b_{\mu},  \cr
w_{\mu} &=& \nabla_\mu \alpha + \zeta \nabla_\mu \beta ,
\end{eqnarray}
where $\varphi$ is a smooth function whose gradient is curl-free, i.e., 
\begin{equation}
(\nabla_{\mu}\nabla_{\nu} -\nabla_{\nu}\nabla_{\mu} ) \varphi =0,
\end{equation}
while the vector field $b_{\mu}$ gives the vorticity, described by its ``field strength" $b_{\mu\nu}$ 
\begin{equation}
\nabla_{\mu}v_{\nu} - \nabla_{\nu}v_{\mu} = \nabla_{\mu}b_{\nu} - \nabla_{\nu}b_{\mu} \equiv b_{\mu\nu}.
\end{equation}
Now both the stress-energy tensor and the mass current should include the contribution of the vorticity. For simplicity, let us first consider the zero temperature $T=0$ cases without $w^\mu$ as in \cite{Lebedev82}
\begin{eqnarray} \label{Tjv}
T^{\mu}_{\nu} &=& -\frac{\partial \mathcal{L}}{\partial v_{\mu}} v_{\nu} - 2 \frac{\partial \mathcal{L}}{\partial b_{\mu\tau}} b_{\nu\tau} + \delta^\mu_\nu \mathcal{L}  \\
j^{\mu} &=&  -\frac{\partial \mathcal{L}}{\partial v_{\mu}} -2 \nabla_\tau  \frac{\partial \mathcal{L}}{\partial b_{\mu\tau}},
\end{eqnarray}
and the conservation law $\nabla_\mu T^{\mu}_{\nu} =0 $ leads to
\begin{equation}
j^\mu b_{\mu\nu} = 0,
\end{equation}
in comparison with the curl-free or irrotational cases in which $b_{\mu\nu} = 0$. Note that $j^{\mu}$ also has a vorticity dependence. Therefore after the decomposition (\ref{new_vw}), besides $b_\mu$ the Lagrangian should also contain $b_{\mu\nu}$ explicitly. From the zero temperature example given in \cite{Lebedev82}, the inclusion of $b_\mu$ and $b_{\mu\nu}$ in the Lagrangian is much more complicated than having a ``kinetic" term $\sim b^{\mu\nu} b_{\mu\nu}$ as one may have imagined, in analogy to the electromagnetic case.  This will be discussed in the next section.

\section{String theory of quantized vorticity}

We give a relativistically covariant description of quantized vorticity in
Minkowski spacetime with metric diag$\left(  -1,1,1,1\right)  $. It is easily
generalized to curved spacetime. The vortex configuration is specified by a
space curve, which sweeps out a world sheet in 4D spacetime. The dynamics of
the vortex line is therefore that of a relativistic string, which has been
widely discussed in the literature \cite{Vilenkin} \cite{Gradwohl}. We
summarize known results from our perspective. 

To begin, we covariantly separate rotational flow from irrotational flow by writing
\begin{equation}
v^{\mu}\equiv\partial^{\mu}\sigma=\partial^{\mu}\chi+b^{\mu}%
\end{equation}
where $\chi$ is a continuous function (whereas $\sigma$ is only continuous
modulo $2\pi)$, and $b^{\mu}$ describes vorticity. We define a ``smooth" order
parameter $\psi,$ with the phase $\chi:$
\begin{equation}
\psi=Fe^{i\chi}%
\end{equation}
The Lagrangian density can be rewritten in terms of $\psi$ and $b^{\mu}:$
\begin{equation}
\mathcal{L}_{0}=\partial^{\mu}\phi^{\ast}\partial_{\mu}\phi+V\left(
\phi^{\ast}\phi\right)  =\left(  \partial^{\mu}+ib^{\mu}\right)  \psi^{\ast
}\left(  \partial_{\mu}-ib_{\mu}\right)  \psi+V\left(  \psi^{\ast}\psi\right)
\end{equation}
This says that we can start with potential flow described by $\psi$, and
introduce vorticity through by introducing a ``gauge field" $b^{\mu}$. The system
is invariant under a local ``gauge transformation" $\psi\rightarrow\psi^{\prime},$
$b_{\mu}\rightarrow b_{\mu}^{\prime}$, with
\begin{align}
\psi^{\prime}  &  =e^{-i\alpha}\psi\nonumber\\
b_{\mu}^{\prime}  &  =b_{\mu}+\partial_{\mu}\alpha
\end{align}
where $\alpha\left(  x\right)  $ is a continuous function; the transformation suggests
an emergent ``gauge symmetry" and is equivalent to a shift $\chi\rightarrow\chi-\alpha$. 
Note that $\bm{b}$ has a dual personality: on the one hand, it is like a magnetic field according
to (\ref{biot-savart}), and on the other hand it is like a gauge field in the present context. Note that $b_\mu$ is constrained by
the vortex quantization condition, which can be represented covariantly as
\begin{equation}
\oint\limits_{C} dx^{\mu}b_{\mu}=2\pi n\text{ \ \ }\left(  n=0,\pm1,2,\ldots\right)
\end{equation}
By means of the Stokes theorem, we can rewrite this as
\begin{equation}
\frac{1}{2}\int_{S}dS^{\mu\nu}b_{\mu\nu}=2\pi n\text{ \ \ }\left(
n=0,\pm1,2,\ldots\right)
\end{equation}
where $S$ is a surface bounded by the closed path $C\,$, $dS^{\mu\nu}$ is a
surface element, and $b^{\mu\nu}$ is the antisymmetric vorticity tensor
defined by
\begin{equation}
b^{\mu\nu}\equiv\partial^{\mu}v^{\nu}-\partial^{\nu}v^{\mu}=\partial^{\mu
}b^{\nu}-\partial^{\nu}b^{\mu}=\left[  \partial^{\mu},\partial^{\nu}\right]
\sigma
\end{equation}
Now we define the ``dual" of vorticity $b_{\mu\nu}$ \cite{Vilenkin}
\begin{equation}
\tilde{b}^{\mu\nu} \equiv \frac{1}{2} \epsilon^{\mu\nu\rho\tau} b_{\rho\tau}
\end{equation}
In terms of the phase $\sigma$, it becomes
\begin{equation} \label{dualvorticity}
\tilde{b}^{\mu\nu} = \frac{1}{2} \epsilon^{\mu\nu\rho\tau} [\partial_\rho, \partial_\tau] \sigma.
\end{equation}
The dual vorticity $\tilde{b}^{\mu\nu}$ is a distribution, e.g. for a static, straight vortex line lying on the $z$-axis, 
\begin{equation}
\tilde{b}^{03} = \frac{1}{2} \delta(x) \delta(y)
\end{equation} 
In general, $\tilde{b}^{\mu\nu}$ perform a projection onto the worldsheet swept by the vortex line. Suppose the worldsheet is parametrized by $x^{\mu} = x^{\mu} (\zeta^a), (a = 0, 1)$, 
$\tilde{b}^{\mu\nu}$ can be written as \cite{Vilenkin}
\begin{equation} \label{projection}
\tilde{b}^{\mu\nu} = \frac{1}{2} \int \delta^{(4)} (x - x(\zeta^a)) d\sigma^{\mu\nu}
\end{equation} 
where $d\sigma^{\mu\nu} \equiv \epsilon^{ab} x^{\mu}_{\, , a} x^{\nu}_{\, , b} d^2 \zeta$ is the area element of the worldsheet.  

These are necessary ingredients connecting the (modified) NLKG to its intrinsic vortex dynamics (or connecting a field theory 
to a ``string" theory). From equations (\ref{dualvorticity}) and (\ref{projection}) it is clear that the vortex dynamics is determined by the phase $\sigma$ and the zero of the modulus $F$.

The vector $b_\mu$ can be related to a Kalb-Ramond potential similar to \cite{Vilenkin}. The decomposition 
\begin{equation} \label{vub}
v_\mu = \partial_\mu \varphi + b_\mu \equiv u_\mu + b_\mu
\end{equation}
with the identification
\begin{equation} \label{b_B}
b_{\mu} = \frac{1}{2} \epsilon_{\mu\nu\lambda\rho} \partial^{\nu} B^{\lambda\rho}
\end{equation} 
where $B^{\mu\nu}$ is the antisymmetric Kalb-Ramond field \cite{Kalb:1974}, generalizes the usual Helmholtz decomposition for a 3-vector ${\bm v}$, 
\begin{equation}
{\bm v} = {\bm v}_{\parallel} + {\bm v}_{\perp}, ~~~\nabla \times {\bm v}_{\parallel}= \nabla \cdot {\bm v}_{\perp} = 0.
\end{equation}
It is easy to see that in equation (\ref{vub}) the ``longitudinal" component $u_\mu = \partial_\mu \varphi$ does not contribute to the vorticity while the ``transverse" component $b_\mu =  1/2 \, \epsilon_{\mu\nu\lambda\rho} \partial^{\nu} B^{\lambda\rho}$ is divergenceless for a regular Kalb-Ramond  field $B^{\mu\nu}$. To see why the relation (\ref{b_B}) is the relativistic generalization of the three-dimensional analogue
\begin{equation}
{\bm v}_{\perp} = \nabla \times {\bm A},
\end{equation}
one may set the components of the Kalb-Ramond field to be ($i, j, k = 1, 2, 3$)
\begin{equation}
B_{i0} = A_i, ~~B_{ij} = \epsilon_{ijk} x^k
\end{equation}
and then obtain
\begin{equation}
b^i =  \epsilon_{ijk}  \partial_j A_k, ~~b^0 = \textrm{const.}
\end{equation}
The relativistic generalization of the three-dimensional vorticity vector 
\begin{equation}
{\bm  \omega} = \nabla \times {\bm v},
\end{equation}
is in a Chern-Simons form \cite{Lebedev82},
\begin{equation} \label{Kmu}
K^{\mu} = \epsilon^{\mu\nu\rho\tau} b_{\nu} \partial_{\rho} b_{\tau}
\end{equation}
whose spatial component contains a term 
\begin{equation}
\epsilon^{i 0 jk} b_{0} \partial_{j} b_{k} = - b_{0} \epsilon^{ijk} \partial_{j} b_{k}. 
\end{equation}
Therefore, the role of the vector potential in ${\bm v} = \nabla \times {\bm A}$ is played by the Kalb-Ramond field $B^{\mu\nu}$ and the role of the vorticity field ${\bm  \omega} = \nabla \times {\bm v}$ is played by the Chern-Simons form 
$K^{\mu} = \epsilon^{\mu\nu\rho\tau} b_{\nu} \partial_{\rho} b_{\tau}$. 

Once we have identified $b_\mu =  1/2 \, \epsilon_{\mu\nu\lambda\rho} \partial^{\nu} B^{\lambda\rho}$, we can use the Kalb-Ramond action \cite{Kalb:1974} as the effective action for a vortex line or a vortex ring. With the Kalb-Ramond field strength
\begin{equation}
H_{\mu\nu\lambda} = \partial_\mu B_{\nu\lambda} + \partial_\nu B_{\lambda\mu} +\partial_\lambda B_{\mu\nu}
\end{equation}
the original nonlinear action for the Klein-Gordon field $\Phi = F e^{i \sigma}$ (at $T=0$) becomes
\begin{equation}
S_0 [F, \sigma] \rightarrow S_0[F, B_{\mu\nu}] = \int d^4 x \bigg[ \partial^{\mu}F \partial_{\mu}F + \frac{1}{6 F^2} H_{\mu\nu\lambda} H^{\mu\nu\lambda} - V(F^2) \bigg] + 2 \pi \int B_{\mu\nu} d\sigma^{\mu\nu}.
\end{equation}
Integrating over the massive $F$ modes for a string solution gives the Kalb-Ramond action \cite{Vilenkin}.

Now we look back at the original Lagrangian and consider its independent variables: We have $v^2 = v^\mu v_\mu$,  $\omega^2 = \omega_{\mu\nu} \omega^{\mu\nu}$ with $\omega_{\mu\nu} \equiv \partial_\mu v_\nu -\partial_\nu v_\mu$, and $h^2 = h_\mu h^\mu $, 
where the Chern-Simon current $h^\mu$ is defined as (similar to Eq. (\ref{Kmu}))
\begin{equation}
h^\mu  = \epsilon^{\mu\nu\rho\sigma} \omega_{\nu\rho} v_\sigma.
\end{equation}
It is easy to see that
\begin{equation}
h^2 = -\frac{1}{2} ( v^2 \omega^2 + 2 v_\mu v^\nu \omega_{\nu\lambda} \omega^{\lambda \mu})
\end{equation}
therefore we see that $v^2, \omega^2, h^2$ can be considered as independent variables in the Lagrangian. One can easily write down other Lorentz invariants such as
\begin{equation}
h^\mu v^\nu \omega_{\mu\nu},~~~h^\mu v^\nu \tilde{\omega}_{\mu\nu}, ~~~\tilde{\omega}_{\mu\nu} \tilde{\omega}^{\mu\nu}, ~~~{\omega}_{\mu\nu} \tilde{\omega}^{\mu\nu}
\end{equation}
where the dual tensor $\tilde{\omega}^{\mu\nu} \equiv \frac{1}{2} \epsilon^{\mu\nu\rho\tau} \omega_{\rho\tau}$.
However, with the help of two identities
\begin{eqnarray}
\omega_{\mu\lambda} \tilde{\omega}^{\lambda \nu} &=& - \frac{1}{4} \delta^{\nu}_{\mu}~ \omega_{\rho\tau}  \tilde{\omega}^{\rho\tau},  \cr
\tilde{\omega}_{\mu\lambda} \tilde{\omega}^{\lambda \nu} &=& \omega_{\mu\lambda} {\omega}^{\lambda \nu} + \frac{1}{2} \delta^\nu_\mu ~ \omega_{\rho\tau} {\omega}^{\rho\tau}, 
\end{eqnarray}
one can show that what is really important is the ${\omega}_{\mu\nu} \tilde{\omega}^{\mu\nu}$ term and the other terms can be reduced to combinations of known terms. Also, note that
\begin{equation}
\omega_{\mu\nu} \tilde{\omega}^{\mu\nu} = \frac{1}{2} \partial_{\mu} h^{\mu}
\end{equation}
similar to the gauge theory cases where $F \tilde{F} \sim \partial_\mu K^{\mu}$, i.e. the topological charge term can be written as the divergence of the Chern-Simons current. Therefore the Lagrangian should only depend on these Lorentz scalars
\begin{equation}
\mathcal{L}_{T=0} = \mathcal{L} \, ( ~v^2, ~\omega^2, ~h^2, ~\omega \tilde{\omega}).
\end{equation}
For the finite temperature cases, $w_\mu$ should be included as well. Neglecting classical vorticity, one can write
\begin{equation}
\mathcal{L}^{T} = \mathcal{L} \, ( ~v^2, ~\omega^2, ~h^2, ~w^2, ~v \cdot w, ~\omega \tilde{\omega}).
\end{equation}
Note that in the present paper we are not intended to write down an explicit Lagrangian for $\mathcal{L}^{T}$. Instead we aimed to show how the relevant degrees of freedom come from the (modified) NLKG/NLSE. In practice what we propose to solve numerically is the original (modified) NLKG/NLSE, similar to what we have done in \cite{Xiong14}.

\section{Magnus force, mutual friction and other forces}

Just like the vorticity is built in the NLKG, forces like the Magnus force, the mutual friction between the quantum vorticity and the normal fluid should be included automatically in the effective potential part of the NLKG. Solving such a NLKG should yield all the effects these forces produce. First let us consider the Magnus force. Any force in vortex dynamics is connected with some velocity by the Magnus relation, connecting the vortex line velocity and the external force per unit length applied to the vortex line (see e.g. \cite{Sonin})
\begin{equation}
{\bm F}= \rho {\bm \kappa} \times ({\bm v}_0 - {\bm v}_L)
\end{equation}
where ${\bm \kappa}$ is the circulation vector of magnitude $\kappa$, ${\bm v}_0$ is the constant velocity that a fluid current passes the vortex line and ${\bm v}_L$ is the velocity of the vortex line. We use a simple example at zero-temperature to demonstrate the existence of the Magnus force. From the decomposition $v_\mu = \partial_\mu \varphi + b_\mu$, we take 
\begin{equation}
\varphi = \omega t, ~~~~\textrm{($\omega$ is a constant)}
\end{equation}
which corresponds to a constant background
\begin{equation}
H^{0}_{ijk} \propto \epsilon_{ijk}
\end{equation}
in terms of the Kalb-Ramond field. With proper gauge the equation of motion of string is \cite{Vilenkin}
\begin{equation}
\mu_0 ( \ddot{q}_\mu - {q''}_{\mu} ) = 4 \pi F_0 (H^{0}_{\mu\nu\lambda} + \cdots) \dot{q}^{\nu} {q'}^{\lambda}
\end{equation}
Its spatial component gives
\begin{equation}
\mu_0 ( \ddot{q}_i - {q''}_i ) = 4 \pi F_0 \epsilon_{ijk} \dot{q}^{j} {q'}^{k}
\end{equation}
Note that $\dot{q}^{j}$ is the vortex line velocity ${\bm v}_L$ and $ {\bm q'}$ is the vortex line tangent. We see that the right-hand side of the above equation can be written as ${\bm v}_L \times {\bm q'}$, which shows the existence of Magnus force in the superfluid rest frame
\begin{equation}
{\bm F}_M \sim {\bm v}_L \times {\bm \kappa}
\end{equation}

Now we consider the mutual friction between quantum vortices and the normal flow. Notice that there is a non-vanishing imaginary part in the effective potential $W$ in Eq. (\ref{Eff_W}}). For simplicity we assume that this imaginary part is a constant $\propto \gamma$. At the non-relativistic limit it reduces to a damped NLSE 
\begin{equation}
( i - \gamma) \hbar \frac{\partial \psi}{\partial t} =  \left( - \frac{\hbar^2}{ 2m }\nabla^2 + \lambda |\psi|^2 - \mu \right)  \psi 
\end{equation}
which for small $\gamma$ becomes approximately \cite{Tsubota13} 
\begin{equation}
i \hbar \frac{\partial \psi}{\partial t} =  ( 1 - i \gamma)  \frac{\delta H[\psi, \psi^*]}{\delta \psi^*} 
\end{equation}
where $H[\psi, \psi^*]$ is the Gross-Pitaevskii energy functional
\begin{equation}
H[\psi, \psi^*] = \int d^3 x \left[ \frac{\hbar^2}{ 2m } |\nabla \psi |^2 - \mu  |\psi |^2 + \frac{\lambda}{2}  |\psi |^4 \right]
\end{equation}
To connect the order parameter $\psi (\bm{x}, t)$ to the motion of the vortex line, we follow \cite{Kawasaki, Nemirovskii2013} and consider $\psi (\bm{x}, t)$ as a functional of the vortex configuration $\psi (\bm{x}, \bm{s}(\xi, t))$. The time-derivative of $\psi (\bm{x}, t)$, for example, is then related to the vortex line velocity $\dot{\bm{s}}$ as
\begin{equation}
\frac{\partial \psi (\bm{x}, t)}{\partial t} = \int_C \frac{\delta \psi (\bm{x}, t)}{\delta \bm{s}(\xi', t)}~ \frac{\partial \bm{s}(\xi', t)}{\partial t} ~ d\xi'
\end{equation}
and then it has been shown in \cite{Nemirovskii2013} that
\begin{equation}
\dot{\bm{s}} = \frac{1 + \gamma^2}{ 1 + \beta^2 \gamma^2} ~\bm{b} + \frac{ \beta \gamma (1 + \gamma^2)}{1+ \beta^2 \gamma^2} ~ \bm{s}' \times \bm{b}
\end{equation}
where the constant $\beta$ is defined in Eq. (\ref{beta}). This allows identifying the mutual friction coefficients \cite{Nemirovskii2013} 
\begin{equation}
\alpha = \frac{ \beta \gamma (1 + \gamma^2)}{ 1 + \beta^2 \gamma^2}, ~~~\alpha' = \frac{(\beta^2 -1)\gamma^2}{ 1 + \beta^2 \gamma^2}
\end{equation}

\section{Conclusions and Discussions}

In this paper we studied relativistic two-fluid model with quantized vorticity via a modified NLKG. An effective potential is introduced to describe the coupling of the superfluid and the normal fluid. It has been shown that such a formulation incorporates vorticity and the related vortex dynamics, and hence, can facilitate numerical analysis which is usually quite complicated from phenomenological point of view (e.g. Schwarz's numerical studies based on vortex filaments).  We also considered the connections to other formulations, especially the duality between scalar field and Kalb-Ramond field,  and the similarity between quantized vortices and global strings. We propose that just like in the zero-temperature pure superfluid cases (as we have shown numerically in \cite{Xiong14}), quantum vorticity and quantum turbulence should be studied using the modified NLKG/NLSE, possibly coupled to other equations depending on the systems or circumstances, in a relativistic or non-relativistic way.

What is the range of validity of the modified NLKG/NLSE? This is a interesting
question, with different answers from different points of view. One can derive
the NLKG/NLSE from the quantum N-body wave function, but the validity of this
approach is limited to weak interparticle interactions, and in the
neighborhood of the ground state of the system, i.e., at low temperatures. The
reason is as follows. First, the assumption that the interparticle potential
is a delta function can be justified only for weak interactions described
through a small S-wave scattering length, which give the equivalent
hard-sphere interaction. Secondly, the derivation from the quantum N-body
problem corresponds to a mean-field approximation, in which one assumes that
effects from excitations from the ground state are small. From this point of
view, then, the NLKG/NLSE is a weak-interaction low-temperature approximation.

In the Ginsburg-Landau theory of phase transitions governed by an order
parameter, on the other hand, the NLKG/NLSE is a purely phenomenological equation,
valid near the transition point of the phase transition. Thus, the order
parameter is assumed to be small. One expands the nonlinear potential in
powers of the order parameters, and just retain the first few terms. From this
point of view, the NLKG/NLSE is valid in the neighborhood of the phase transition
that creates the order parameter. Which view one adopts would depend on the application.

\appendix

\section{Notations on phenomenology of vortex filaments}

We follow Schwarz's formulation \cite{Schwarz}.
The vortex line is represented by a space curve with its position described by ${\bm S}(\xi, t)$, where $\xi$ parametrizes the curve and $t$ is the time. The curl-free condition is violated on this curve,
\begin{equation}
{\bm \omega} = \nabla \times {\bm v} = \kappa \int d{\bm S} ~ \delta ({\bm r} -  {\bm S}(\xi, t)),
\end{equation}
where the integration is along the vortex line  ${\bm S}(\xi, t)$. At distance relatively far from the vortex line, the above equation and the condition $\nabla \cdot {\bm v} =0$ yield a Biot-Savart type equation
\begin{equation}
{\bm v}_{\textrm{\tiny{ind}}}  = \frac{\kappa}{4 \pi} \int d\xi' \frac{[{\bm S}(\xi', t)- {\bm S}(\xi, t)] \times {\bm S'}_{\xi'}}{|{\bm S}(\xi', t) - {\bm S}(\xi, t)|^3}
\end{equation}
where ${\bm S}' =\partial {\bm S}/\partial \xi $.  The local superfluid velocity ${\bm v}_{sl} $ can be affected by an external flow, more precisely ${\bm v}_0$, the superfluid velocity at large distance from any vortex line \cite{Donnelly}
\begin{equation}
 {\bm v}_{sl} = {\bm v}_{\textrm{\tiny{ind}}} + {\bm v}_0.
\end{equation}
The Magnus force is given by 
\begin{equation}
{\bm f}_{M} = \rho_s \kappa \frac{{\bm S}'}{|{\bm S}'|} \times ( {\bm v}_L - {\bm v}_{sl})
\end{equation}
where ${\bm v}_L= \dot{{\bm S}} \equiv d{\bm S}/dt $ is the velocity of the vortex line. The Magnus force reflects the difference between vortex line velocity, ${\bm v}_L$, and the local superfluid velocity, ${\bm v}_{sl}$. The next factor determining the vortex line dynamics is the mutual friction between the quantum vortices and the normal component of the superfluid. The mutual friction ${\bm f}_{D} $ acting on a unit length of the vortex line is
\begin{equation}
{\bm f}_{D} = D_1  \frac{{\bm S}'}{|{\bm S}'|} \times \bigg[  \frac{{\bm S}'}{|{\bm S}'|} \times ( {\bm v}'_{n} - \dot{{\bm S}}) \bigg] + D_2  \frac{{\bm S}'}{|{\bm S}'|} \times ( {\bm v}'_{n} - \dot{{\bm S}})
\end{equation}

The vortex line motion is described by Schwarz's equation \cite{Schwarz}
\begin{equation}
\dot{{\bm S}} = {\bm v}_{\textrm{\tiny{ind}}} + {\bm v}_0 + \alpha   \frac{{\bm S}'}{|{\bm S}'|} \times ( {\bm v}_{ns} - {\bm v}_{\textrm{\tiny{ind}}}) - \alpha'  \frac{{\bm S}'}{|{\bm S}'|} \times \bigg[  \frac{{\bm S}'}{|{\bm S}'|} \times ( {\bm v}_{ns} - {\bm v}_{\textrm{\tiny{ind}}}) \bigg]
\end{equation}
where ${\bm v}_{ns} \equiv {\bm v}_{n} - {\bm v}_{s}$ is difference between the average normal-fluid velocity and the applied superflow field, and the coefficients $\alpha, \alpha'$ can be expressed in terms of the coefficients $D_1, D_2$. This is the basic equation for describing problems on the motion of the vortex lines, in particular, the {\it vortex tangle} problem \cite{Schwarz}.

We summarize the notations for vortex dynamics as follows,
\begin{itemize}
\item ${\bm S}(\xi, t)$ --- position of the vortex line. ($\dot{{\bm S}} \equiv d{\bm S}/dt,  {\bm S}' \equiv \partial {\bm S}/\partial \xi, \cdots $);
\item ${\bm v}_{n}$ --- effective, or macroscopically averaged normal fluid velocity \cite{Schwarz};
\item ${\bm v}_{s} $--- macroscopically averaged superfluid velocity;
\item ${\bm v}_{L}$ --- vortex line velocity, ${\bm v}_{L} = \dot{{\bm S}}$;
\item ${\bm v}_{\textrm{\tiny{ind}}}$ --- the velocity of superflow induced by the curvature of the vortex line;
\item ${\bm v}_{sl}$ --- local superfluid velocity, ${\bm v}_{sl} = {\bm v}_{\textrm{\tiny{ind}}} + {\bm v}_s$;
\item ${\bm f}_M$ --- the Magnus force, ${\bm f}_{M} = \rho_s \kappa \frac{{\bm S}'}{|{\bm S}'|} \times ( {\bm v}_L - {\bm v}_{sl})$;
\item  ${\bm f}_D$ --- the mutual friction has different expressions, e.g. \cite{Schwarz, Donnelly} \\ 
${\bm f}_{D} = - \alpha \rho_s \kappa \frac{{\bm S}'}{|{\bm S}'|} \times \bigg[  \frac{{\bm S}'}{|{\bm S}'|} \times ( {\bm v}_{n} - {\bm v}_{sl}) \bigg] - \alpha' \rho_s \kappa \frac{{\bm S}'}{|{\bm S}'|} \times ( {\bm v}_{n} - {\bm  v}_{sl})$
\end{itemize}

\end{document}